**Qumus: Realization of An Embodied AI Quantum Material Experimentalist**

Lihan Shi[1,2,#], Zhaoyi Joy Zheng[1,2,#], Xinzhe Juan[3,#], Yimin Wang[3,#], Ming Yin[2,4,#], Mayank Sengupta[1,4], Kristina Wolinski[1], Yanyu Jia[1], Jingzhi Shi[5], Derek Saucedo[1,6], Neill Saggi[1], Haosen Guan[1], Kenji Watanabe[7], Takashi Taniguchi[8], Ali Yazdani[1], Mengdi Wang[2,4], Sanfeng Wu[1,*]

[1] Department of Physics, Princeton University, Princeton, New Jersey 08544, USA
[2] Department of Electrical and Computer Engineering, Princeton University, Princeton, New Jersey 08544, USA
[3] Department of Computer Science and Engineering, University of Michigan, Ann Arbor, Michigan 48109, USA
[4] Princeton AI Lab, Princeton University, Princeton, New Jersey 08544, USA
[5] Currently unaffiliated, Incoming graduate student at Princeton ECE
[6] Department of Physics and Astronomy, California State University, Northridge, Northridge, California 91330, USA
[7] Research Center for Electronic and Optical Materials, National Institute for Materials Science, 1-1 Namiki, Tsukuba 305-0044, Japan
[8] Research Center for Materials Nanoarchitectonics, National Institute for Materials Science, 1-1 Namiki, Tsukuba 305-0044, Japan
[#] These authors contributed equally to this work
[*]Email: sanfengw@princeton.edu

**While modern Large Language Models[1–4] (LLMs) and agentic artificial intelligence (AI)[5–9] have demonstrated transformative capabilities in digital domains, the realization of embodied AI capable of real-world scientific discovery[10–18] remains a difficult frontier. The advancements are hindered by the inherent complexity of integrating high-level reasoning, multimodal information processing and real-time physical execution. Here we introduce Qumus, the first AI quantum materials experimentalist. Physically embodied within a robotic mini-laboratory, Qumus is an intelligent, multimodal, and multi-agent system designed for the creation and nano-processing of atomically thin two-dimensional (2D) materials and stacked van der Waals (vdW) structures. Qumus autonomously navigates the full scientific cycle, from hypothesis generation and protocol planning to multi-step experimental execution, result analysis and reporting, acting as an experimentalist. Markedly, the system has achieved, for the first time, the AI-creation of graphene, as well as the first AI-fabrication of complex nanodevices including atomically thin field-effect transistors via vdW stacking. Qumus excels at these tasks by demonstrating autonomous error correction and closed-loop experimentation. Our results establish a generalizable framework for self-improving embodied AI systems that learn directly from the quantum world, opening a pathway toward accelerated discovery in quantum materials, electronics and beyond.**

## Main

2D quantum materials are a class of atomically thin crystals and their derived structures that exhibit rich electronic, optical, and quantum phenomena. The field was sparked by the invention of the Scotch-tape-based mechanical exfoliation of graphene in 2004[19–21], which launched extensive worldwide endeavor to investigate 2D crystals. In principle, thousands of known layered crystals can potentially be exfoliated down to the atomically thin limit[22,23], which can be further integrated into van der Waals (vdW) stacks and moiré materials that can substantially alter the material properties[24–31]. In sharp contrast to the enormous possibilities, experiments on 2D materials and moiré materials in the laboratory remain largely concentrated on a small set of familiar crystals, such as the graphene class and several transition-metal dichalcogenides, despite the extensive efforts over two decades[24–31]. Beyond the lab, many promised applications, including vdW-material-based electronics[24,28,30,32–34], have yet to be achieved. A central bottleneck in both laboratory and industrial-scale applications is the inefficient manual process of synthesis, identification, transfer, and assembly of high-quality flakes and devices[35–44]. The many-step workflows are time-consuming, difficult to reproduce, and heavily dependent on expert intuition and iterative troubleshooting; for many air-sensitive materials, repeated manual handling is impractical for human researchers.

To address these challenges, previous efforts have automated certain steps of 2D material workflows[45–51] using either rule-based approaches or machine learning methods from the pre-LLM era. While such methods have made important progress, they fall short of providing a seamless, end-to-end autonomous process spanning exfoliation, flake search, transfer, and vdW stacking without human intervention. More importantly, these systems, not driven by language models, lacked the cognitive framework required for scientific inquiry. Lacking the ability to reason, hypothesize, or iterate, they functioned as automated tools rather than genuine AI experimentalists. In this work, we introduce Qumus, the first closed-loop AI quantum material experimentalist designed for the fully autonomous creation of 2D crystals, vdW stacks, and devices. Qumus is a self-evolving, multimodal, and multi-agent AI system embodied within a robotic minilab.

## Qumus AI Architecture

A truly intelligent AI experimentalist should possess a suite of defining capabilities within a closed-loop framework: interpreting scientific requests, reasoning, proposing hypotheses, planning and executing experiments, observing and analyzing results, updating plans as needed and reporting the outcome. Together, these functions form an integrated and iterative feedback cycle (**Fig. 1a**) capable of achieving complex scientific objectives. Inspired by the structure of human research groups, where a principal investigator (lead agent) manages multiple researchers (sub-agents) responsible for distinct sub-tasks, Qumus employs an organizational structure where central AI orchestrates a team of sub-agents with specialized Skill sets (**Figs. 1b & c, Extended Data Table 1**) to ensure efficient collaboration and communication.

The lead agent, Qumus, plays the central role of reasoning, planning, coordinating teamwork and communicating with external users (e.g., humans). It interprets the scientific goal, determines whether the request is sufficiently specified, translates the goal into one or more experimental milestones, makes an experimental plan, and executes it by delegating tasks to sub-agents. The lead agent also integrates feedback reported by the specialized agents, reformulates the task when prerequisite subtasks are required, selects among candidate experimental proposals, and generates final reports summarizing workflows, parameters, results, and evaluation.

Key sub-agents include Project Manager Agent, Lab Manager Agent, Device Expert Agent and Processing Agent, and their functional modules are summarized in **Fig. 1b** and **Methods**. We outline their respective roles. The *Project Manager* has access to prior experimental knowledge, including literature, local experimental history, and registered Skills and recipes to inform planning. The *Lab Manager* assesses laboratory readiness by checking material inventory, instrument status, and tool positioning via computer vision. The *Device Expert* generates and refines device-layout proposals, including 2D flake stacking order, positions, and twist angles given the available materials and target device specifications. The *Processing Agent* converts requested experimental goals into executable laboratory actions: it retrieves existing workflows or self-generates new ones, and reports outcomes back to Qumus and the database for future reuse.

A key design principle of the Processing Agent is that it operates via a hierarchical workflow structure consisting of (i) *"Atom Workflows"*, which are fixed unit executable functions provided by the user at the instrumental level to perform simple operations, such as robotic arm control, stage movement control, camera control, vacuum control, tape handling, etc; (ii) *"Molecule Workflows"*, which are combinations of multiple atomic workflows (building blocks) that together accomplish a more complex function, such as chip exfoliation, automated search, or flake re-location; and (iii) *"Assembly Workflows"*, which are larger combinations of atomic and molecular workflows that implement extended multi-stage experimental procedures, such as repeated exfoliation, stacking, or other longer fabrication sequences. This hierarchical structure is key to achieving robust experiments by fixing the "atom workflows" while enabling AI self-generated workflows at the "Molecule" and "Assembly" levels, important for Qumus self-evolution.

**The Robotic Minilab**

We designed and built a robotic hardware platform that integrates automated Scotch-tape exfoliation, optical inspection, precision 2D crystal transfer and vdW stacking under a microscope and sample storage into a single compact workstation (**Fig. 1d**). The system combines a tape-transfer gantry, robotic handling arms, heated sample stages, a microscope inspection module, and multiple precision motion stages, collectively enabling full hardware automation of every physical step in 2D material mechanical exfoliation, processing and stacked device fabrication. Tape handling, including dispensing, positioning, peeling, and collection of used tape, is automated through a dual-wheel motorized feed system. Materials, such as raw material carriers, bare silicon chips, and device substrates, are handled and transported by two robotic arms positioned around the central workspace (exfoliation and stacking stages) and among three material storage tables.

Qumus is equipped with multimodal computer vision across scales, from the macroscopic to the microscopic. To recognize all macroscopic objects necessary for the functions, we train a computer vision algorithm using YOLO[52–54] using two top-mounted cameras (**Extended Data Fig. 1)**. We attach micro-QR codes to key objects to ensure robust AI recognition, accurate coordinate mapping, and sample/consumable information to be processed. At the microscopic scale, 2D flake identification has been well explored. In our setup, optical inspection is handled by a multi-magnification optical microscope equipped with a motorized focus module for automated z-axis focusing, enabling autonomous microscale on-chip feature detection and 2D flake identification across large substrates[55]. Temperature-controlled stations with vacuum holders immobilize the substrates and stamps during exfoliation, search, and stacking operations. Submicron-level alignment and transfer are achieved by two coordinated precision motion systems: a multi-axis stamp alignment stage and a multi-axis sample positioning stage, which together provide the translational and rotational degrees of freedom required for high precision 2D materials stacking. More details about the robotic hardware can be found in **Methods** and **Extended Data Fig. 2**.

## The First AI Creation of Graphene

With Qumus AI embodied within the robotic minilab, the system can interact with human researchers (or external AI agents) using natural language, analyze tasks, plan, and execute experiments. We first demonstrate that Qumus can autonomously and robustly plan and complete complex experimental workflows. A user can simply enter a user request through the user interface (**Methods**), e.g., "Can you give me a graphene flake?" **Fig. 2a** illustrates the information flow of the collaborative agentic system upon receiving the request. In a typical process, Qumus first consults the Project Manager to understand the nature of the experiments and receives suggested experimental recipes, based on which it proposes and broadcasts a plan book (**Fig. 2b**). Qumus then executes the plan, step by step, by delegating the task in each step to a corresponding sub-agent and collects feedback after each step. For instance, Lab Manager may be triggered to provide information about the inventory of raw materials and instrumental status, and if no graphene or unsearched graphene chip is found in the inventory, Processing Agent will be triggered to execute the full exfoliation and flake-search process (**Figs. 2c & d, Extended Data Fig. 3**) until a piece of graphene is created, found, recorded and reported. **Fig. 2e** shows a representative graphene flake created by Qumus. The entire process is fully autonomous, with multiple closed-loop decision-making processes by Qumus and its managed agents without any human input (**Supplementary Demo Video 1**). The only required human involvement is to provide raw materials and electricity.

## Characterizations of Qumus as an Experimentalist

The multi-agent, collaborative architecture of Qumus ensures robust end-to-end experimental execution. We characterized Qumus across 6 leading LLMs: GPT, Gemini, Claude, Grok, Qwen and DeepSeek, all of which successfully accomplished the requested experiments. While each model performs rather consistently within itself, Qumus exhibits distinct behavioral characteristics across models, analogous to the differing personalities of human experimentalists. As an example, by decomposing each experiment into reasoning, observation, and execution phases (**Fig. 2f**), we can quantify metrics such as "bias for action" that capture each Qumus-LLM model's tendency toward deliberation versus direct execution (**Methods**). We characterize the personalities along seven dimensions: protocol alignment, caution, bias for action, token efficiency, agent efficiency, consistency and report quality, as summarized in radar charts in **Fig. 2g** & **Extended Data Fig. 4.**

The defining characteristic of an AI experimentalist is its ability to reason, hypothesize, design experiments to test the hypotheses, observe, analyze, and autonomously iterate this cycle toward a scientific goal or conclusion. To test this ability, we posed an open-ended user prompt "I want a graphene flake larger than 200 $\mu m^2$", and cleared all prior experimental history, forcing Qumus to begin its exploration from scratch. This triggered Qumus to explore the parameter space for exfoliating a large graphene flake: substrate heating temperature ($T$), heating dwell time ($t$), number of massage cycles ($n$), and tape peel-off speed ($v$). **Fig. 3a** showcases a representative run of Qumus powered by Claude Sonnet 4.6. Much like a seasoned human experimentalist, Qumus began by formulating a starting hypothesized recipe based on common sense and general knowledge, executed the experiment, and then iteratively refined its approach based on observed outcomes. In this case study, the initial parameter guess was already a strong start (likely reflecting the existing knowledge in the training data of LLM) but produced a near-miss result. Crucially, Qumus responded to the outcome with a highly logical, observation and evidence-based reasoning process, systematically exploring the parameter space and eventually reverting toward more productive recipes when new recipes yielded significantly lower flake yield. After five consecutive experimental runs spanning over four hours, Qumus successfully achieved the goal and produced a graphene flake larger than 200 $um^2$ without any human intervention.

The demonstration establishes Qumus as an experimentalist as defined in **Fig. 1a**. It also demonstrates that Qumus is goal-oriented, equipped with long-term memory and capable of autonomously executing complex, multi-step scientific tasks over extended time horizons.

**Autonomous Error Correction & Closed-Loop Experimentation**

We further highlight interesting features of Qumus. A particularly notable capability is its ability to detect and correct errors autonomously, crucial for operational robustness in real-world lab settings. **Fig. 3b** and **Supplementary Demo Video 2** summarize a case study of Qumus creating hexagonal boron nitride (hBN), an important dielectric material for 2D devices. First, a totally unexpected error was introduced mid-experiment: a human removed its newly exfoliated silicon chip that was actively under processing, without any human notification to the system. Qumus autonomously detected the problem through computer vision, verified the absence of the chip, and then formulated a recovery plan to resolve the issue (i.e., re-exfoliating a new chip). In this new run, however, the Processing Agent hallucinated, an intrinsic LLM error, and mistakenly labeled the material under processing as "Graphene" instead of "hBN". This again resulted in a null outcome. Qumus analyzed the situation and again came up with a solution to exfoliate a new chip, eventually successfully completing the experiment. The entire error-detection and correction process required no external input or human intervention of any kind. It demonstrates that Qumus is a closed-loop AI experimentalist with robust performance capable of autonomous error correction.

**Qumus Orchestration and AI-Creation of an Atomically Thin Transistor**

To demonstrate Qumus' ability to execute complex, multi-stage experimental projects, a user asked the system: "Can you give me a graphene transistor?" A $SiO_2$/conductive Si substrate with pre-deposited metal contacts was loaded as an input material registered in the database. **Fig. 4a** summarizes the resulting agentic orchestration and cooperation network. Qumus first analyzed the request and consulted the Project Manager to define an executable device strategy and determined the device stack as hBN/graphene placed on metal contacts on $SiO_2$, with conductive Si substrate serving as a back gate. It then decomposed the objective into sequential subtasks and delegated each step to the appropriate specialist agent (**Fig. 4b**). In this case study (**Figs. 4c-g**), the Lab Manager confirmed that graphene flakes and a user-loaded conductive $SiO_2$/Si substrate with prepatterned electrodes were already present in the material database. However, no hexagonal boron nitride (hBN) flake was available, and Qumus therefore delegated the Processing Agent to acquire an hBN flake. After hBN flakes were exfoliated, searched, and stored in the materials database, Qumus delegated the device layout to the Device Expert, which selected the best candidate flakes and designed the device layout based on these selected flakes under geometric and physical constraints, returning the selected flake information, stacking order and device layout. Qumus then passed the approved design and complete stacking payload to the Processing Agent, which autonomously controlled the full transfer module to pick up and stack the flakes onto the electrode substrate via 2D-material dry transfer. The full process took approximately 90 minutes and comprised 30 steps and 18 decision-making calls, all successfully completed by Qumus and its collaborative agent network without human intervention (**Supplementary Demo Video 3**). **Figs. 4h & i** show the resulting Qumus-fabricated graphene field-effect transistor, demonstrating a marked achievement for AI design and fabrication of a quantum material device. More details about the robotic hardware can be found in **Methods** and **Extended Data Fig. 5**.

**Self-Evolving Modules**

Qumus is intrinsically self-evolving[9]. This is evidenced by its ability to plan experiments based on prior history, made possible by a set of structured databases that record not only materials created, but also prior experimental traces across all levels including reasoning chains, individual experiments, and overarching projects. Beyond memories, Qumus progressively accumulates experimental skills, including composite workflows and fabrication recipes (both can be self-generated) that were tested in real experimental runs. In subsequent experiments, the Project Manager can retrieve and inform relevant Skill sets to Qumus (**Figs. 1b & c**), enabling self-improvement. Here we have only reported the initial experiments conducted by Qumus. We expect Qumus soon to expand its capabilities to a broader suite of materials and devices, with iterative enhancements to its performance.

## Discussion and Outlook

Qumus is highly versatile and readily applicable to the vast library of 2D materials beyond those demonstrated here. Its integration into an inert-atmosphere glovebox environment is a critical next step, as it will enable the autonomous processing of air-sensitive materials, broadening the accessible chemical and structural space. We anticipate that Qumus, as an AI experimentalist, will act as a significant catalyst, accelerating the pace of discovery in quantum materials, vdW nanostructures and devices, including the experimentation on those remaining unexplored due to the limitations of manual processes. The primary bottlenecks currently constraining our system are instrumental rather than algorithmic. As noted in **Fig. 2f**, the total time cost is dominated by hardware-level latencies, such as mechanical actuation, optical focusing, and thermal equilibration times. These constraints will be substantially improved by next-generation hardware optimized for high-speed operations. Concurrently, the rapid evolution of AI systems in foundational language models, memory-augmented agents, multi-modal sensing and perception will further enhance Qumus's decision-making and self-evolving capabilities.

Qumus marks the emergence of the first AI quantum experimentalist, providing an unprecedented interface for AI to engage directly with the quantum mechanical aspects of our world. Looking forward, the architecture demonstrated here serves as a modular blueprint for a new class of embodied AI experimentalists. We envision this framework being adapted across a wide spectrum of scientific instrumentation and laboratories, from material synthesis to advanced characterizations[56–60]. Ultimately, Qumus could function as a key node within a larger, collaborative agentic network, where decentralized AI agents synchronize efforts to accelerate large-scale scientific exploration.

## Acknowledgement

This work is supported by the Materials Research Science and Engineering Center (MRSEC) program of the National Science Foundation (DMR-2011750) and the Gordon and Betty Moore Foundation through Grant GBMF11946. S.W. acknowledges support from AFOSR through awards number FA9550-23-1-0140 and FA9550-25-1-0354, and support from the Princeton Laboratory for Artificial Intelligence and the Princeton IP Accelerator Fund. A.Y. is supported by U.S. Army Research Office MURI project under grant number W911NF-21-2-0147 and W911NF261A052 as well as NSF grant DMR-2312311, and NSF grant OMA-2326767. K.W. is supported by the Department of Defense (DoD) through the National Defense Science & Engineering Graduate (NDSEG) Fellowship Program. D.S. is supported by the NSF Research Experiences for Undergraduates (REU) program at Princeton University through the NSF MRSEC program mentioned. K.W. and T.T. acknowledge support from the CREST (JPMJCR24A5), JST and World Premier International Research Center Initiative (WPI), MEXT, Japan.

## Author Contributions

S.W. conceived and designed the project. L.S. and Y.J. led the development of AI-compatible compact hardware, assisted by Z.J.Z. and H. G. Z.J.Z., X.J., L.S., and Y.W. led the development of the AI architecture and prompts, assisted by M.Y. and J.S. Y.M., M.S., K.W., D.S., and N.S. led the development of computer vision algorithms, assisted by L.S. and Z.J.Z. K. W. and T. T. provided hBN crystals. S.W., M.W., and A.Y. co-supervised the project. S.W., Z.JZ., and L.S. wrote the paper with input from all authors.

**Data Availability**

The data that support the findings of this study are available from the corresponding author upon request.

**Code Availability**

Code developed and used in this work will be available via Github at [link to be provided].

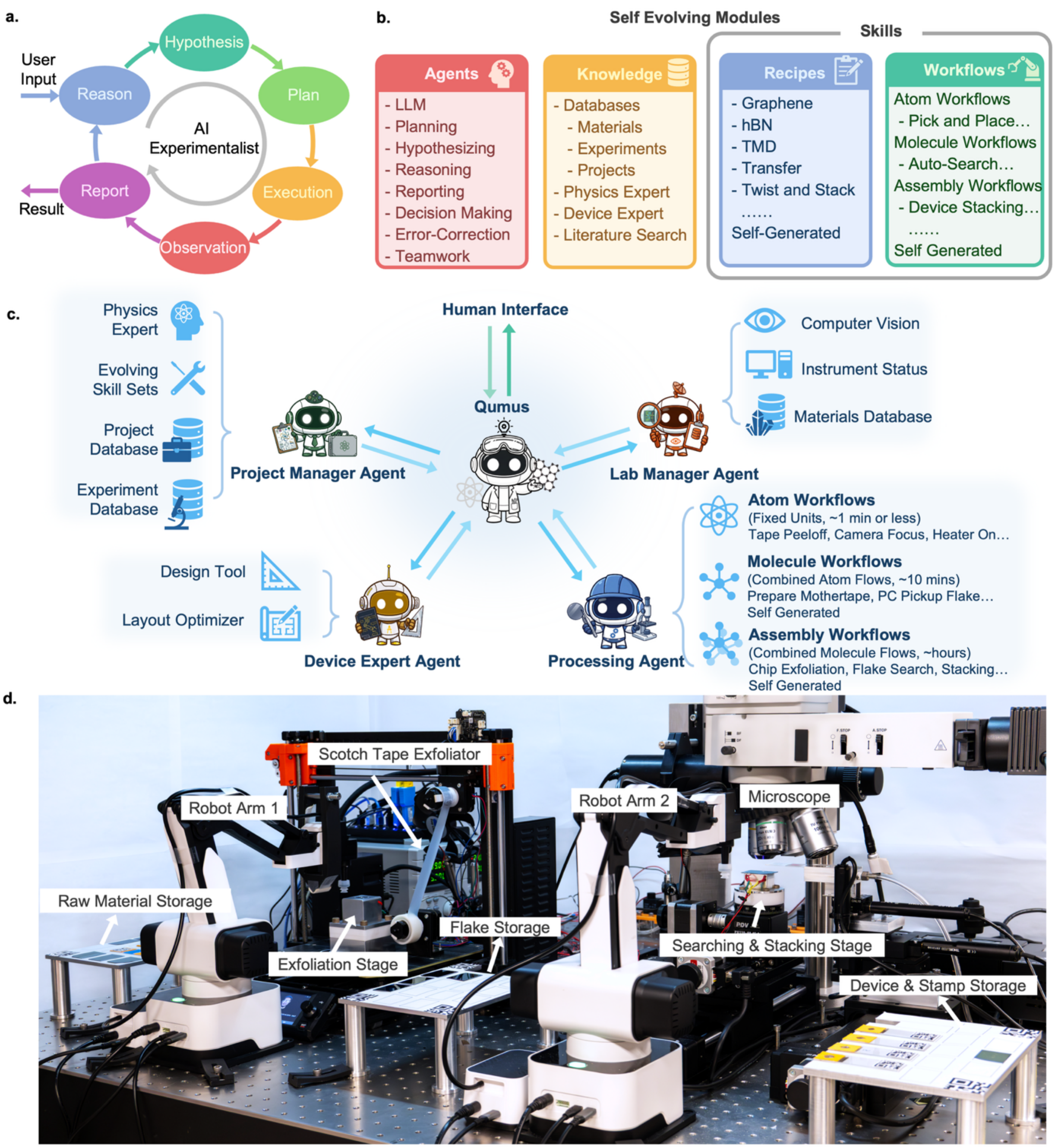

**Fig. 1. Qumus AI architecture and fully robotic minilab. a,** Defining characteristics of an AI experimentalist. **b**, Key self-evolving modules of Qumus, including LLM-agents, memory and knowledge systems, and skills including instrumental workflows and materials/devices realization recipes. **c,** Qumus architecture for efficient multi-agent collaboration and robust performance. **d,** A compact, fully robotic minilab consisting of vacuum- and temperature-controlled stages for 2D material mechanical exfoliation, optical flake search, flake transfer and stacking, along with robotic arms, storage modules, cameras, and microscope systems.

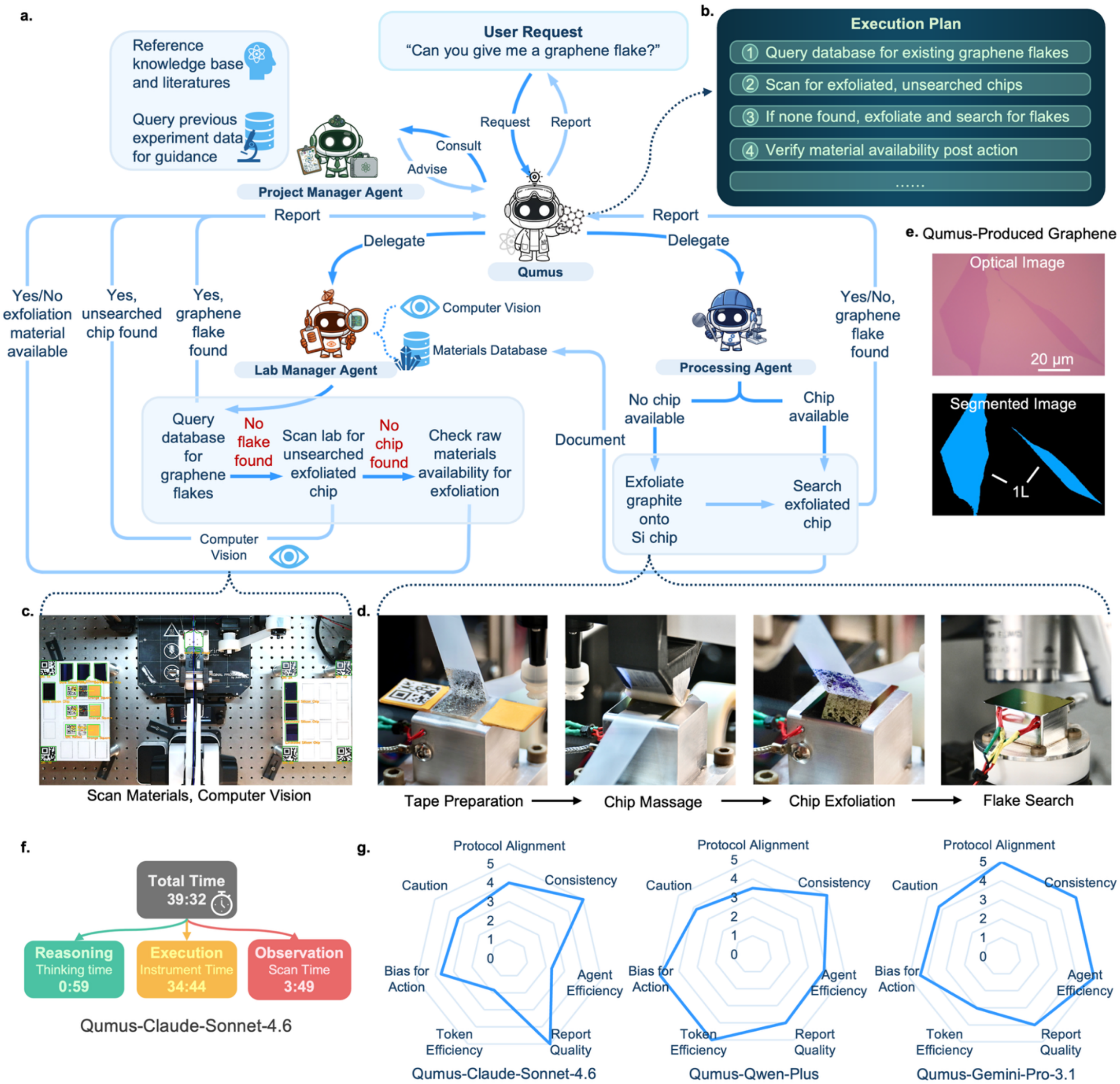

**Fig. 2. The first AI creation of 2D quantum materials and Qumus personality. a**, A representative organizational response of Qumus and information flow chart, upon receiving a user request of a graphene flake. **b**, A representative plan book initiated by Qumus. **c**, Computer vision for identifying key objects during mechanical exfoliation. **d**, A series of instrumental performance realizing mechanical exfoliation and flake search under the optical microscope. **e**, A representative graphene flake created by Qumus. Top: Optical Image; bottom: corresponding segmentation. **f**, An example of total time and its distribution over reasoning, execution and observation, for completing the entire process of providing a graphene flake to users. The major time is instrumental execution, as expected. The reasoning and observation times may vary depending on the chosen LLM. **g**, Selected radar charts characterizing the different "personalities" of Qumus when powered by different LLMs.

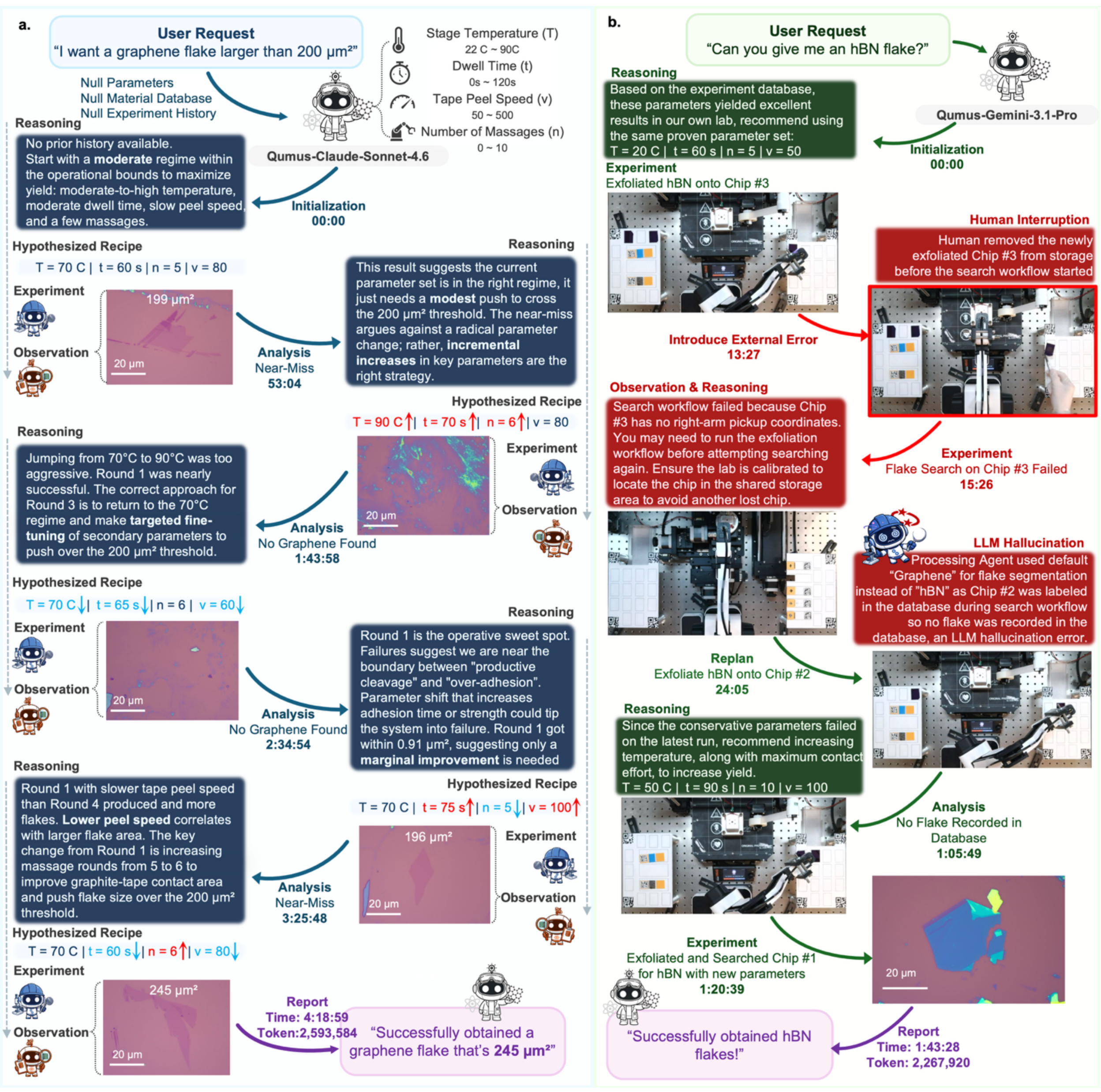

**Fig. 3. Qumus behaves as a closed-loop experimentalist and corrects errors autonomously. a,** Summary of the performance and workflows of Qumus-Claude-Sonnet-4.6, upon receiving a user request. It shows that Qumus can iteratively complete the full process of reasoning, hypothesizing, planning, executing, observing, and analyzing to achieve a goal, with each loop testing a new set of hypothetical parameters based on observations from previous experiments. No human intervention during the entire process. **b**, Qumus-Gemini-Pro-3's response to two errors. First, a human interruption removed the chip under active processing during an intermediate step, causing the workflow to fail. Qumus diagnosed the error and replanned on a new chip. A subsequent LLM hallucination incorrectly segmented flakes using the wrong material label. Qumus recognized no hBN flake is recorded in the database and succeeded by exfoliation and searching a new chip with updated parameters, producing hBN flakes as requested, without human intervention or input.

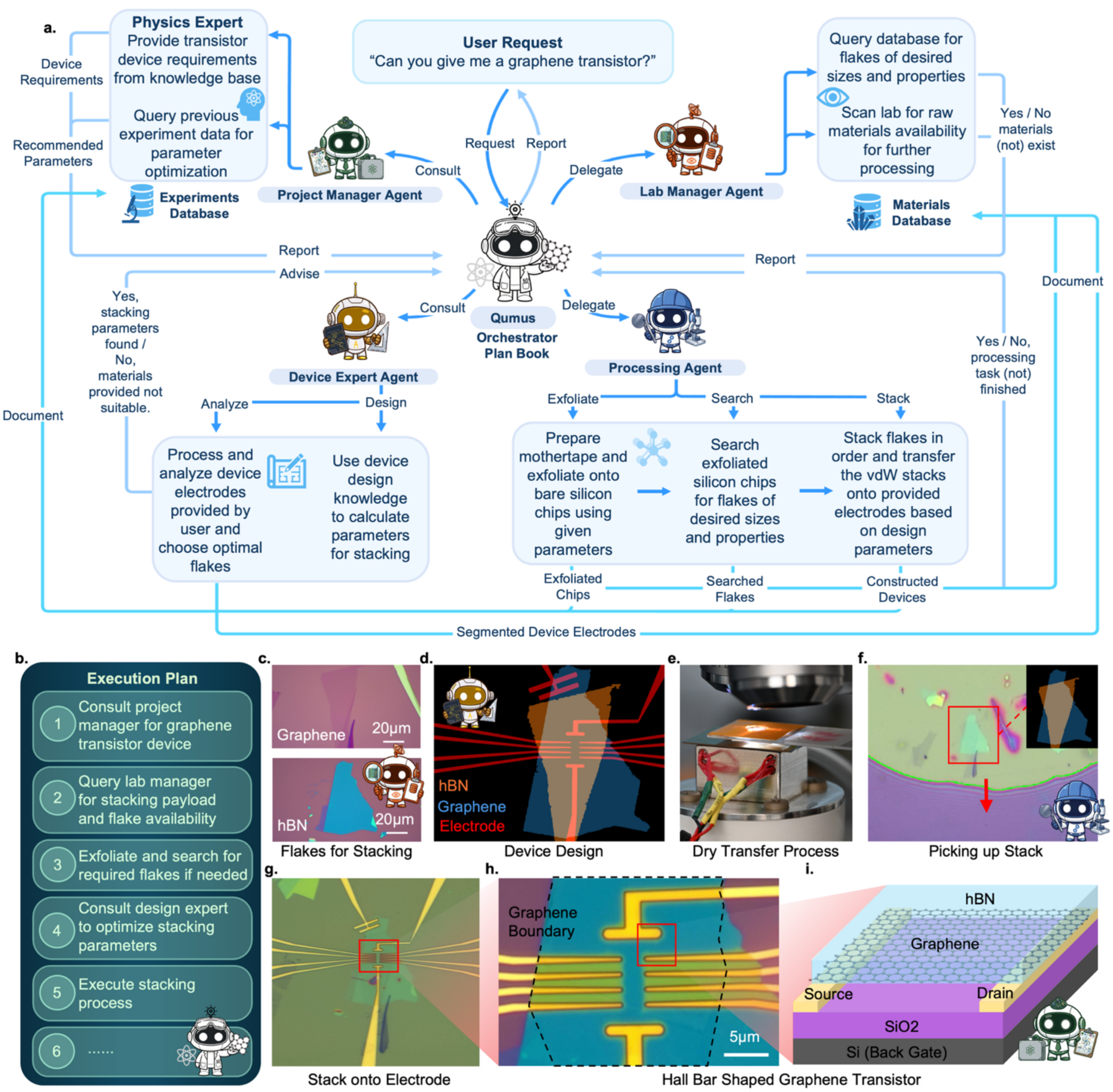

**Fig. 4. Multi-agent orchestration & autonomous construction of complex devices. a,** Workflow and information flow chart of Qumus orchestrating agents to complete the user request of constructing a graphene transistor device. **b,** Initial plan book generated by Qumus. **c,** Optical images of graphene and hBN flakes produced by the Processing Agent and stored in the Materials Database. **d,** Design layout created by the Device Expert using the user-provided metal electrodes, selected graphene and hBN flakes that meet the transistor device requirement. **e,** Image showing the Processing Agent manipulating the stamp to perform flake transfer using the 2D dry-transfer technique. **f,** Optical image showing the polymer film contacting the substrate (yellow area) and picking up the stack, with hBN aligned over graphene according to the design layout in **d**. **g,** Optical image showing the polymer film placing the stack onto the bottom electrodes, following the design layout in **d**. **h,** Optical image at 100x magnification of the final stack after dissolving top PC layer. **i,** Schematic illustrating the graphene field-effect transistor.

## Methods

### Robotic Hardware

The Qumus hardware system is a compact, fully automated platform for 2D material exfoliation, flake identification, transfer, and device fabrication. The hardware consists of a tape-exfoliator module implemented based on a 3D-printer arm, three materials storage tables, two robotic handling arms, temperature-controllable stages with vacuum chucks, a motorized optical microscope with automated focus control, a multi-axis sample positioning stage, and a multi-axis stamp alignment stage. The left side of the workstation is used for exfoliation operations, and the right side is used for microscope flake search, and transfer/stacking operations for device construction. Two overhead cameras monitor the macroscopic workspace and provide visual logging for the presence and fiducial locations of raw material carriers, silicon chips (classified as bare silicon chips, exfoliated chips, and device chips), tapes, and stamps **(Extended Data Fig. 2)**.

All hardware operations are registered with Qumus through hardware-specific software controllers. The controllers include robot-arm, relay, motor, power-supply, sample-stage, transfer-stage, microscope, focus, turret, camera and temperature-control interfaces. These controllers are wrapped into fixed executable tools used by the Processing Agent. This setup allows the agents to choose among valid operations and parameter ranges, while the deterministic controller layer performs the low-level serial, motion, and acquisition commands.

### Agent Prompts and Skills

Qumus is implemented as a ReAct-style tool-calling LLM-agent system in which a lead agent coordinates specialized sub-agents through structured task messages and receives textual or JSON-like reports. Each agent is initialized with a role-specific prompt defining its responsibilities, tools, constraints, output format, and handoff protocol **(Extended Data Fig. 6)**. The lead Qumus agent serves as the global orchestrator: it interprets user requests, plans steps, delegates tasks, incorporates returned observations, and revises plans when feedback invalidates earlier assumptions. Sub-agents do not communicate freely; information flows through Qumus or shared database records, keeping delegation traceable and responsibility explicit. Each specialized agent is equipped with role-specific **Skills**, where a skill is a reusable capability consisted of tools, workflows, and recipes of parameters for processing. These skills range from simple database queries to multi-step experimental workflows, including exfoliation recipes in which parameters such as peeling speed, contact force, and temperature etc., are optimized based on the material, substrate, and thickness requirements **(Extended Data Table 1)**.

Communication uses structured templates specifying the task, required data or physical objective, constraints, and expected deliverable. Sub-agents return concise reports with observations, records, artifacts, execution status, or recommended actions, which Qumus converts into later steps or final summaries. Safety is enforced through prompt rules and tool restrictions: agents must stay within role, avoid and reject unsupported assumptions, and use handoff-ready formats, while database tools are read-only and hardware tools can execute only predefined workflows within specified operational limits.

**Role Summaries of Sub-Agents**

*(a) Project Manager Agent* supports project-level planning based on prior knowledge including literature and local experimental history relevant to the requested scientific task. The supporting modules for Project Manager include (i) Physics Expert Agent, that provides scientific guidance regarding materials, device concepts, physical constraints, and relevant external or online knowledge sources; (ii) Project Database, which stores higher-level project records, project-level goals, summaries, and outcomes; (iii) Experiment Database, which stores detailed process-level records including workflows, execution histories and outcomes (both successful and failed); and (iv) Skill Sets, which are records of optimized workflows and experimental recipes based on prior experience.

*(b) Lab Manager Agent* evaluates laboratory readiness and resource availability. It determines whether suitable flakes, raw materials, and other consumables are available, and verifies whether the hardware system is ready to run. It owns supporting modules including (i) Materials Database, which stores material inventory and flake-level information; (ii) Instrument Status Monitor, which checks the status of instruments and system components; and (iii) Computer Vision YOLO **(Extended Data Fig. 1)**, which verifies tool/material position and availability.

*(c) Device Expert Agent* generates device-layout proposals based on the requested device function and available candidate materials, determining the stacking order and spatial configuration of 2D flakes. It operates through modules including (i) Flake Selector, which selects compatible flakes based on geometric coverage, and (ii) Layout Designer, which determines the rotation and translation in of each material (represented as 2D mask) by enforcing geometric constraints such as contact, overlap, and exclusion relationships.

*(d) Processing Agent* executes hardware workflows to complete requested experimental goals. Upon receiving a fabrication or characterization task from Qumus, it determines whether a suitable prior workflow already exists, executes that workflow if available, or self-generates a new workflow if no suitable prior workflow exists. It is responsible for carrying out mechanical exfoliation, flake search, stacking, and related fabrication tasks, and reports execution history, selected parameters, and outcomes back to Qumus and databases for future reuse and learning.

**Qumus Personality Characterization**

We characterized model-dependent Qumus behavior across seven dimensions that capture reliability, efficiency, bias for action, caution, protocol alignment and reporting quality. For each model, metrics were computed from complete autonomous experimental runs with identical Qumus system prompts and user input ("can you give me a graphene flake?"). The analysis is based on recorded agent traces, tool calls, timing logs and final reports. For each quantitative metric, raw values were converted to a normalized five-point score to enable comparison across models. The best-performing model for a given metric was assigned a score of 5, and all other models were scored relative to that best value. Consistency was defined as the fraction of successful runs among all attempted 5 runs for each Qumum-LLM system. Agent Efficiency was defined as the inverse of the total time spent in non-execution agent activity, including reasoning and observation. Token Efficiency was defined as the inverse of the total number of input and output tokens used during a run. Bias for Action was defined as the fraction of total agent steps that involved concrete hardware execution or delegation of an executable action. Caution was defined as the fraction of total agent steps devoted to reasoning, observation, verification or readiness checking before action. Protocol alignment was scored qualitatively on a five-point scale according to how closely the agent's plan, delegation choices and execution order followed the intended protocol specified in the agent prompts. Report quality was also scored qualitatively on a five-point scale according to factual accuracy, completeness, clarity and usefulness of the final experimental summary **(Extended Data Fig. 4)**.

## Macroscopic Computer Vision and YOLO Training

Qumus uses a trained computer-vision module to monitor the macroscopic state of the physical lab, localizing chips, tapes, glass slides, stages, stamps, and storage-table fiducials before and after robotic actions. Because objects can be placed at arbitrary in-plane orientations, detection is formulated as instance segmentation: each object is returned as a polygon mask, from which centroid, size, and in-plane angle are extracted for robotic alignment. QR codes on key objects provide robust labels and coordinate anchors in addition to learned class labels **(Extended Data Fig. 1)**.

*Model*: We fine-tuned Ultralytics YOLOv8 segmentation models from COCO-pretrained checkpoints. Training images were collected directly from two overhead RGB cameras, each streaming 1920 × 1080 frames at 30 fps, to capture realistic lighting, reflections, and robot-arm occlusions. Separate datasets were collected for the exfoliation and stacking workspaces, covering object classes including silicon chips, scotch-tape crystals, glass-slide QR codes, orange square fiducials, exfoliation stages, transfer stages, stamp holders, and table QR codes. After augmentation, the exfoliation dataset contained 261 images and the stacking dataset contained 100 images.

*Training*: Models were trained in PyTorch on Apple Silicon using the MPS backend for 30 epochs at 1024 × 1024 resolution, batch size 2, SGD with cosine learning-rate decay, a 3-epoch warm-up, and early stopping on validation mAP.

*Deployment and Post Processing*: During deployment, each frame is passed through YOLOv8x with a confidence threshold of 0.5. Each retained polygon mask is converted into a pose descriptor ($x$, $y$, $w$, $h$, $\theta$), where ($x$, $y$) is the mask centroid, $w$ and $h$ are edge lengths, and $\theta$ is the short-edge orientation. The angle is wrapped to [−90°, +90°] and sign-flipped to match the right-handed stage convention. Glass-slide QR detections are paired with orange-square detections using relative angle and separation to compute slide orientation. Final detections, masks, confidences, labels, and pose descriptors are serialized to JSON and written atomically to both rolling latest files and per-frame archives for real-time use by the Lab Manager and Processing Agent.

**Microscopic Computer Vision and Segmentation**

Qumus uses an explainable rule-based algorithm to segment and classify nanomaterial flakes during flake search. This approach generalizes across materials with limited training data, often requiring fewer than five labeled images. The pipeline consists of four steps: LAB conversion, Edge Detection, LAB equalization, and Mask Classification **(Extended Data Fig. 3)**.

*LAB conversion*: The RGB microscope image is converted to CIE-LAB color space, where Euclidean color distances better approximate human-perceived color differences.

*Edge Detection*: Edge detection is then performed independently on each LAB channel using multiple dilation-erosion kernels with different window radii and iteration counts. Large-radius kernels capture low-contrast edges from ultrathin graphene or hBN, while smaller kernels produce sharper edges for higher-contrast flakes and reduce sensitivity to dirt or debris. The resulting edge masks are inverted and combined to segment flakes across a wide range of optical contrasts.

*LAB Equalization*: Images are then LAB-equalized to correct uneven illumination. The algorithm identifies the chip background by comparing large-area masks to a preset standard background LAB color. It samples 2000 random background points and interpolates between them to estimate the illumination field of a flake-free chip. This field is subtracted from the original LAB image and replaced with the standard background color, producing a normalized image with spatially uniform lighting.

*Mask Classification*: each segmented mask is classified using benchmark LAB points learned during training. The average LAB color of each mask is compared to these benchmarks by Euclidean distance, assigning labels such as monolayer, bilayer, trilayer, four-or-more layers, bulk, or debris for graphite and transition metal dichalcogenides, and thickness labels for hBN flakes.

**Chip Corner Detection and Flake Re-Localization**

To enable rapid and reliable re-localization of previously identified flakes, we establish a chip-based coordinate system for each sample. Specifically, the bottom-left corner of the chip is defined as the origin, and the position of each flake is recorded as a vector relative to this reference point. This representation allows the system to avoid exhaustive re-scan of the entire chip during subsequent operations. Instead, the target flake can be directly located by combining the reference corner position with the relative rotation and translation of the chip, enabling efficient and precise flake re-finding **(Extended Data Fig. 5)**.

*Chip Corner Detection*: The chip is segmented from the background using a combination of intensity and color information to ensure robustness under varying imaging conditions. RGB cues are used to resolve ambiguities when grayscale contrast is insufficient. After the chip is placed on the sample stage, live image acquisition starts, coupled with real-time stage jogging movement, allowing continuous detection of the chip region. The background image segmentation mask is used to extract the chip edge, from which corner candidates are identified based on local geometry. The bottom-left corner is used as a reference, and its image coordinates are recorded together with a reference optical image of the corner region, enabling reliable matching in subsequent operations.

*Flake re-localization*: After the chip containing the target flake is placed back onto the sample stage, the reference corner is first re-identified using the procedure described above to establish the chip's current position. The detected corner region is then matched to the stored reference image using feature-based registration, from which a planar homography is estimated. The homography encodes the relative rotation and translation of the chip between the two observations. The rotation angle is extracted from the transformation and used to map the original flake position, recorded as a vector relative to the reference corner, into the current coordinate frame. The transfer stage is subsequently moved according to the transformed coordinates to bring the target flake into the field of view.

**Stacking Process**

*Flake Alignment*: After coarse re-localization, fine alignment is performed to precisely position the target flake. A central region from the original reference image (taken at 20x magnification from the initial auto-search workflow) is matched to the current field of view using normalized cross-correlation. To account for in-plane rotation, the reference image is rotated over a range of candidate angles, and a coarse-to-fine search is used to determine the optimal alignment. Edge-based flake segmentation masks (instead of optical images) are used to eliminate illumination variations. The resulting rotation and translation are used to update the stage position for precise flake alignment.

*Contact Detection*: Two key steps are involved during stacking: detection of the initial contact point between the PC film and the substrate via Newton's rings, and monitoring of the resulting contact area to ensure full coverage of the target flake. As the transfer stage approaches in the out-of-plane direction, images are continuously compared to a reference frame using image differencing. The resulting difference image is normalized and a threshold is applied to identify regions of intensity changes associated with contact formation. Connected-component analysis with area and shape filtering is used to extract a valid contact region. The first detected region indicates the contact point, while the largest region is subsequently tracked to monitor the expansion of the contact area. Its contour is used to verify that the target flake is fully covered with a predefined safety margin. This visual feedback is coupled with real-time motion of the transfer arm, enabling closed-loop control of autonomous transfer.

*Digital Map and Flake Matching*: A digital map of flakes within the field of view is constructed by enhancing the image and extracting flake regions using edge detection, morphological operations, and connected-component analysis. The resulting mask defines the spatial distribution of flakes. At selected stages during transfer, the target flake is identified by matching a reference mask within the current digital map using template matching with scale and rotation variation. The detected position is used to assess the flake state and verify the transfer outcome.

**Device Layout Design**

Device layouts are generated by Design Agent from experimentally acquired flake masks and electrode geometries using a flake selector and a layout designer**.**

*Flake Selector*: Material and electrode images are first converted into binary masks, with electrodes further separated into functional regions. Candidate flakes are then evaluated based on geometric compatibility. For example, top encapsulation layers, such as hBN, are selected by computationally testing different rotations and translations of candidate flakes to identify the one that provides maximum coverage of the target material, and only candidates that sufficiently cover the underlying flake are retained. Among these, flakes with larger usable area are preferred.

*Layout designer:* Device requirements are expressed as geometric constraints between materials and electrodes, including contact, overlap, and exclusion relations. The layout is represented as a pixel-wise binary map, allowing efficient evaluation of interlayer interactions. Materials are first coarsely aligned using centroid matching and principal-axis orientation to establish the required contacts. The configuration is then refined by optimizing the rotation and translation of each material using a global search (differential evolution) to maximize constraint satisfaction. The optimized configuration defines the final device layout and stacking parameters.

**Fabrication Methods**

*Raw Material Carrier Preparation:* To supply the Qumus robotic lab with two-dimensional material crystals for exfoliation, searching, and stacking, we prepared raw-material carriers using glass slides. Crystals of 2D materials, including hBN and graphite, were placed on 0.75-inch-wide tape, which was mounted at the center of each glass slide using double-sided tape. Micro-QR codes encoding the material type were attached to one side of each glass slide to enable computer-vision-based identification and material labeling. Orange square markers were attached to the opposite side of the glass slide to improve reliability during robotic-arm pickup.

*2D Flake Transfer:* The transistor devices were assembled using a standard 2D dry transfer method. Qumus-produced hBN and graphene flakes were sequentially picked up with a PC/PDMS polymer stamp under optical microscope alignment, forming hBN/Graphene/electrodes stacks. The stacks were transferred and placed onto $SiO_2$/Si wafers with prepatterned Au metal electrode.

*Metal Contacts Preparation:* We used electron-beam lithography to define contact regions on $SiO_2$/conductive Si substrate, then we performed cold tank development, reactive ion etching, and electron-beam metal deposition to deposit metal contacts.

**Materials, Experiments and Project Databases**

Qumus stores persistent states in three SQLite databases with distinct ownership boundaries. The Materials Database records physical inventory, optical images, flakes, chips, stamps, raw-material carriers, electrodes and devices. The Project Database records hardware topology, instrument status, workflow templates, workflow parameters, workflow runs, step logs and artifacts. The Experiment Database records user sessions, agent runs, nested agent calls, reasoning steps, tool invocations, tool responses and generated artifacts. The database split allows the Lab Manager, Processing Agent, Design Agent and Project Manager to share information while preserving clear responsibility for each type of state.

The database tables are relational. Important relationships include flake-to-material, flake-to-coordinate, flake-to-chip, flake-to-image, electrode-to-chip, device-to-chip, chip-to-slot, chip-to-corner-image, workflow-run-to-template, workflow-step-to-run, agent-step-to-agent-run and tool-response-to-tool-call links. These relations allow a final experimental outcome to be traced back to the material, chip, image, workflow, agent decision and tool output that produced it.

**UI and Server**

Qumus has a browser-based UI and FastAPI backend that let users control the system, monitor agents, view live lab state, inspect databases, review designs, analyze devices, and audit experiment history. The homepage is the main control surface. Users send natural-language requests to the lead agent, select agents, watch real-time reasoning/tool-call streams, follow workflow plans, and stop runs if needed. The backend launches each agent run as an isolated subprocess rather than running agents inside the web server. This keeps long-running autonomous experiments and hardware-control dependencies separate from the UI server. Agent output is streamed to the browser using server-sent events. Structured events are parsed into reasoning logs, active-operation updates, plan steps, tool calls, delegation events, images, and final responses. The UI provides live MJPEG camera streams during robotic operations using the two overhead RGB cameras and camera streaming avoids contention with the vision system.

The Database page exposes capped, read-only views of the materials, processes, and experiments SQLite databases, allowing users to inspect lab state without browser-side write access. The Interactive Design page supports user inspection and manual correction of generated stack designs, material/electrode masks, device-segmentation outputs. The Settings page controls model identifiers, max step limits, verbosity, hardware ports, camera indices, and API settings, allowing the same UI to work with different LLM providers and physical setups.

**Open-Source Software**

Qumus acknowledges the following open-source resources: smolagents, LiteLLM, Ultralytics YOLO, PyTorch, OpenCV, NumPy, SciPy scikit-image, BoofCV, PyBoof, SQLite, SQLAlchemy, FastAPI, Uvicorn, pySerial, pythonnet and Arduino IDE.

**Supplementary Demo Videos are available at [https://qumus.ai](https://qumus.ai)**

## Extended Data

**a.** **Real Lab View (Left Side)** **b.** **Real Lab View (Right Side)**

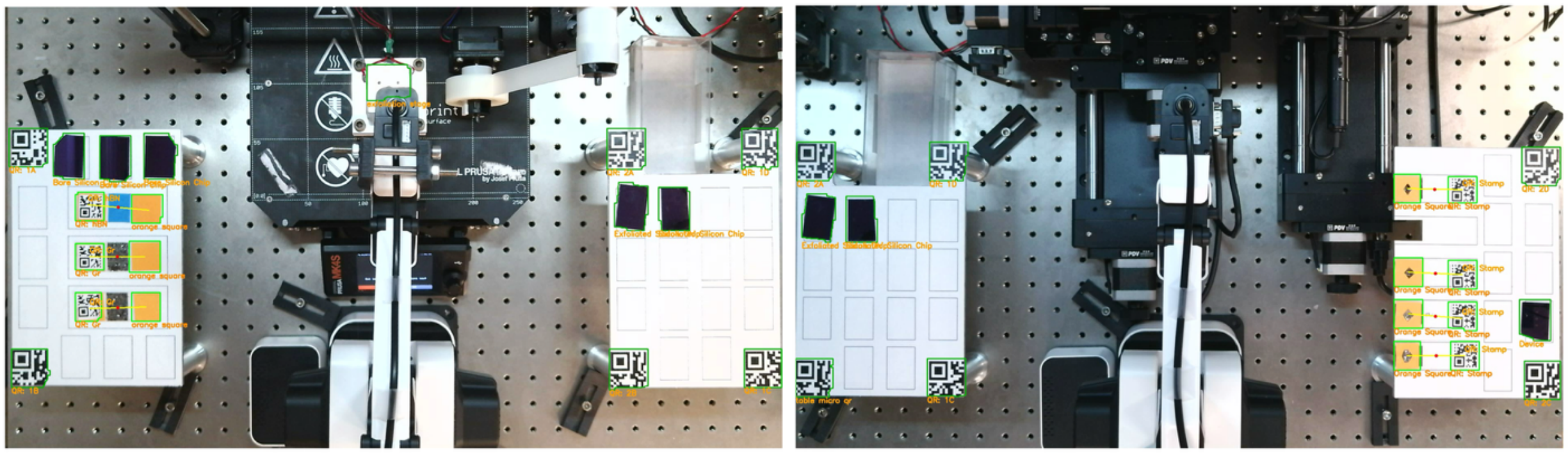

**c.** **Consolidated Digital Lab View**

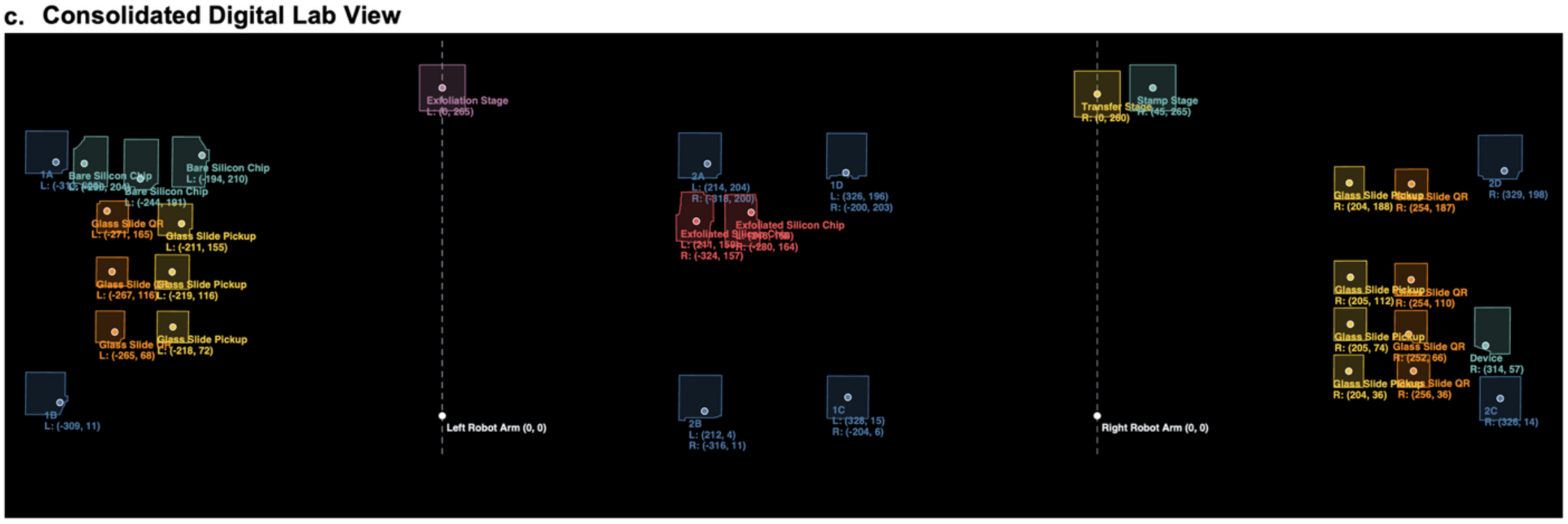

**Extended Data Figure 1 | Macroscopic Computer Vision. a,** Real camera view of the left-side lab space with detected objects, QR-code messages, and oriented bounding boxes overlaid. The left-side view is used primarily for exfoliation operations, including the inventory of raw-material carrier, bare silicon chip, and exfoliated chips, the status of the scotch tape, and the occupancy of the exfoliation stage. **b,** Real camera view of the right-side lab space. The right-side view is used primarily for search and stacking operations, including exfoliated chip, stamp, and device storage. **c,** Consolidated digital lab view generated from the two camera streams after object detection, QR-code identification and coordinate transformation. Detected objects are mapped into a common robot / table coordinate frame and labelled by object type and position, providing the Lab Manager Agent and Processing Agent with a shared representation of current workspace state for inventory synchronization, readiness verification, and material handling coordinates.

a.

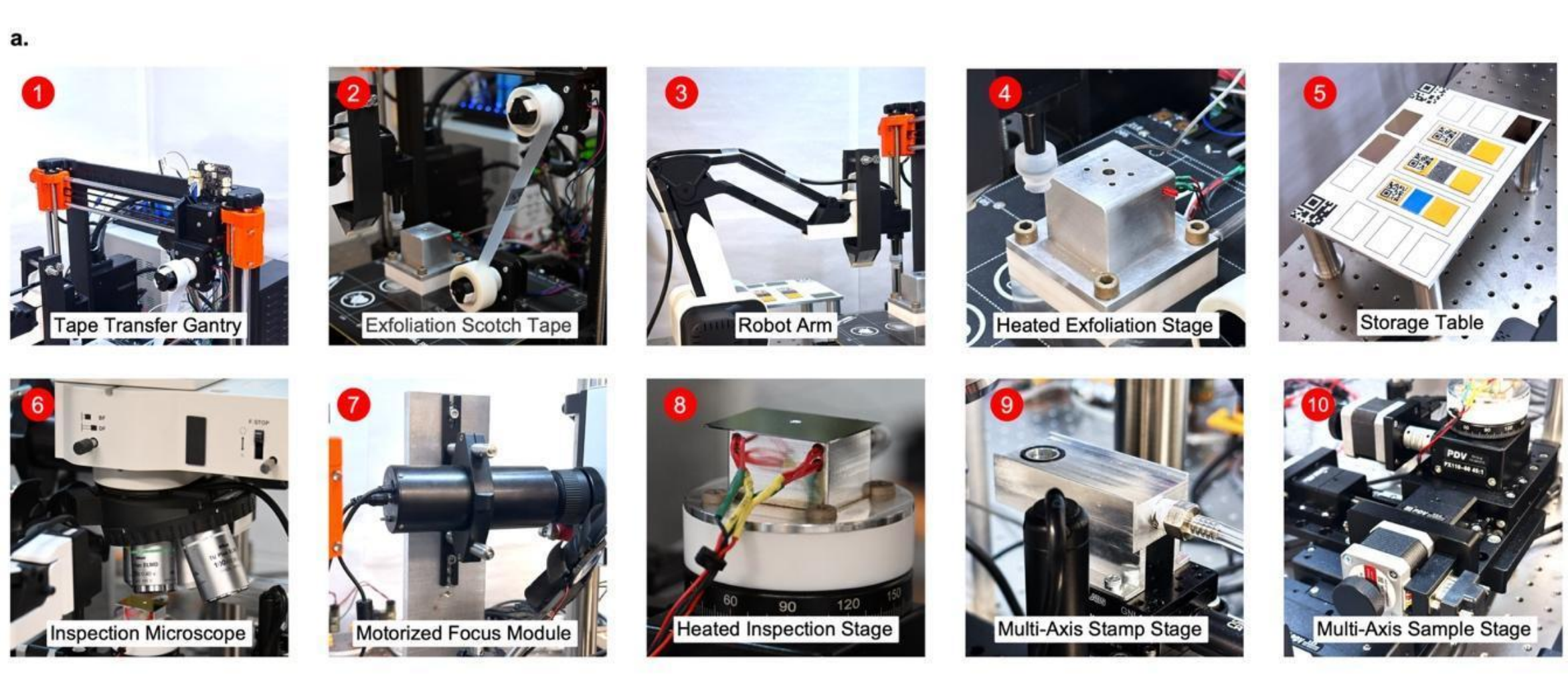

b.

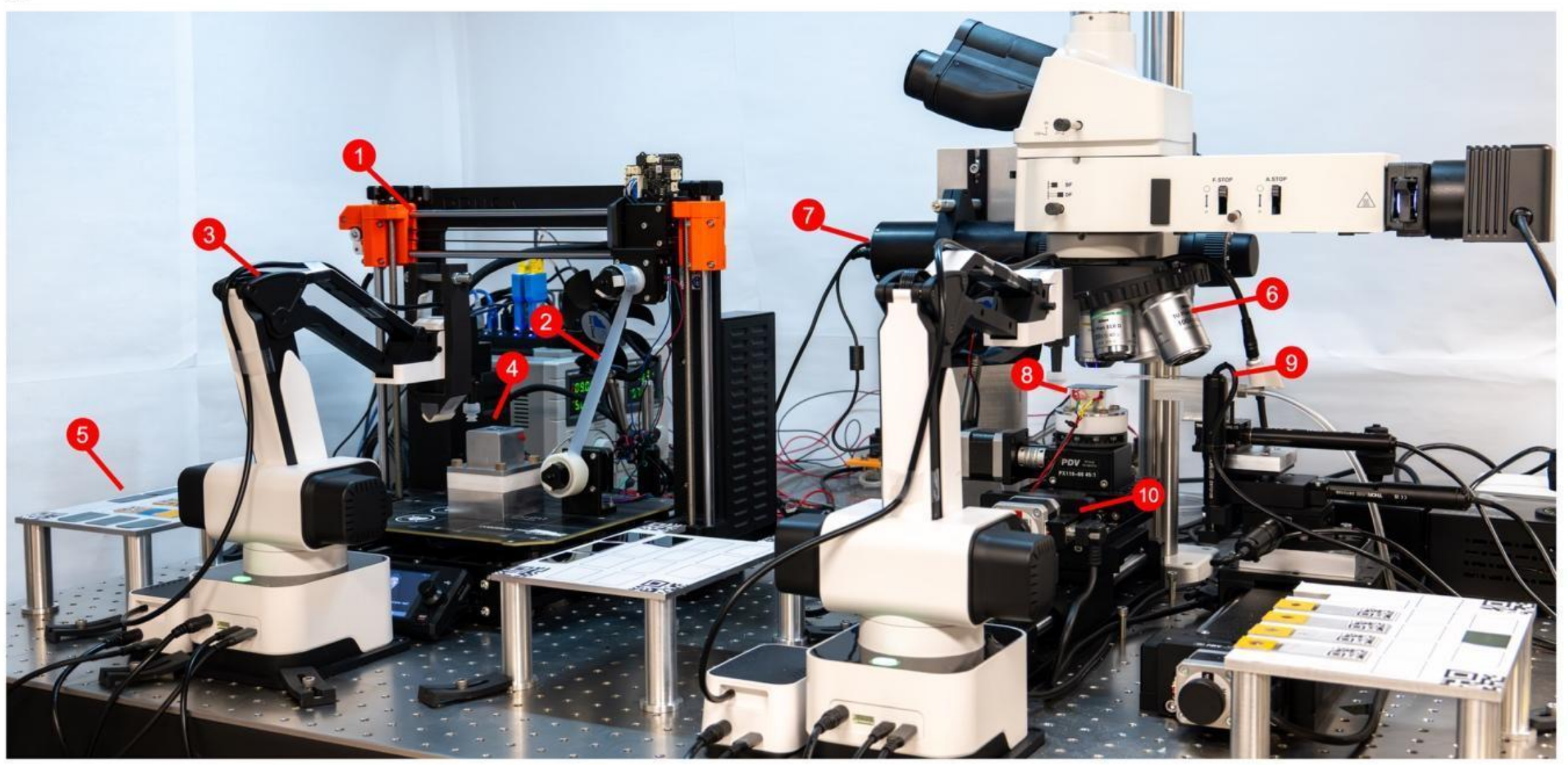

**Extended Data Figure 2 | Labeled hardware components of the automated exfoliation and transfer platform. a,** Close-up view of 10 hardware components labeled. **b,** Assembly view of the hardware system

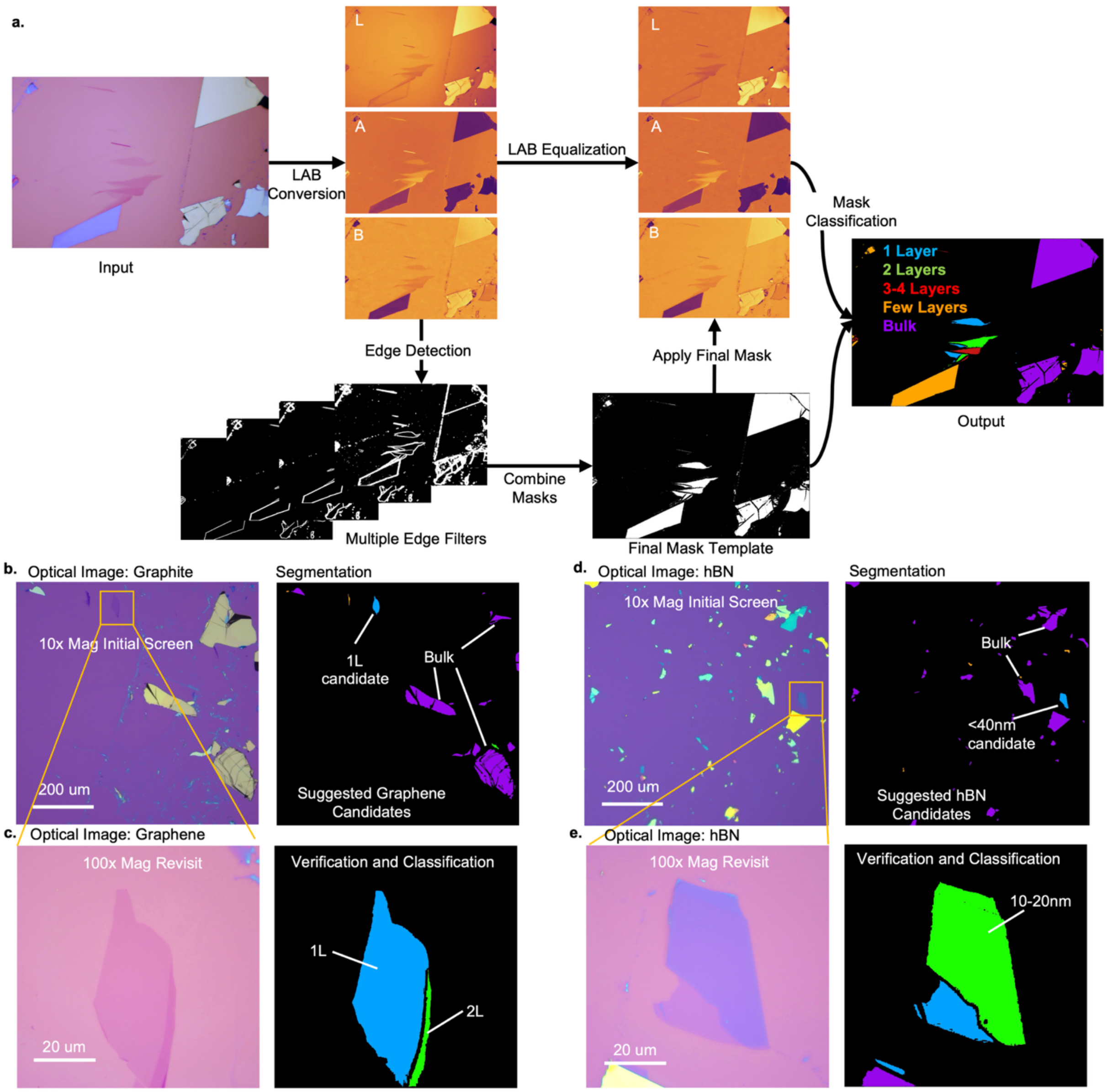

**Extended Data Figure 3 | Flake Segmentation and Representative Results. a,** Workflow of flake segmentation and detection, consisting of LAB conversion, edge detection, LAB equalization, and mask classification. Each connected region is classified by comparing its mean LAB color to material-specific benchmark colors, producing various thickness labels. **b**, Low-magnification (10x) graphite optical image and the segmentation result showing candidate flakes, where colors denote estimated layer number: blue: monolayer; green: bilayer; red: trilayer; orange: few-layer; purple: bulk. **c,** High-magnification (100x) optical image to verify and classify the thickness label of candidate flakes identified in (b), using the same graphene layer color code as in (b). **d**, Low-magnification (10x) optical image of hBN and segmentation result showing candidate flakes, colors represent estimated thickness: blue: 0-10nm; green: 10-20nm; red: 20-30nm orange: 30-40nm; purple: over 40nm. **e**, High-magnification (100x) optical image to verify and classify the thickness of candidate flakes identified in (d), using the same hBN layer color code as in (d).

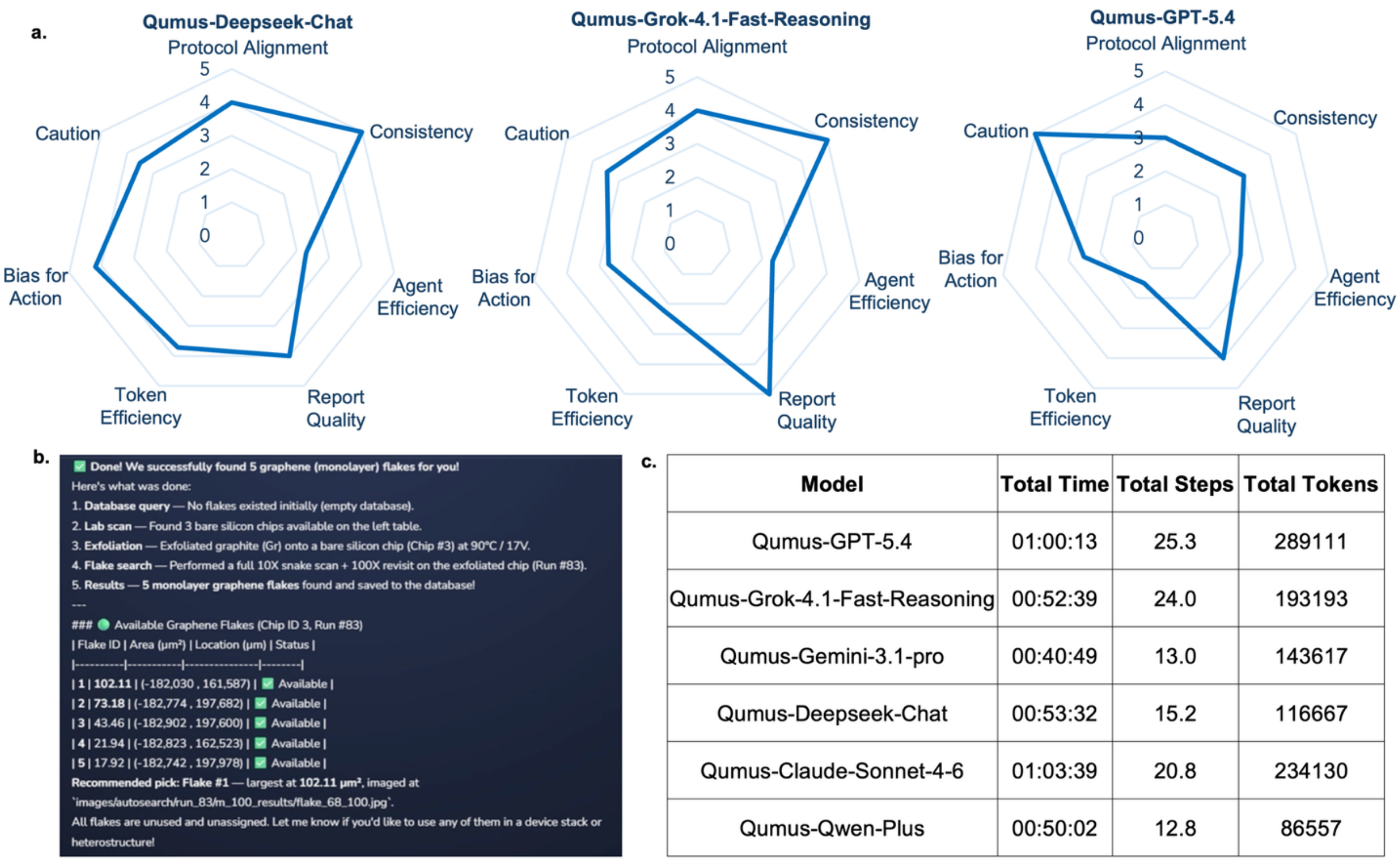

| Model | Total Time | Total Steps | Total Tokens |
|---|---|---|---|
| Qumus-GPT-5.4 | 01:00:13 | 25.3 | 289111 |
| Qumus-Grok-4.1-Fast-Reasoning | 00:52:39 | 24.0 | 193193 |
| Qumus-Gemini-3.1-pro | 00:40:49 | 13.0 | 143617 |
| Qumus-Deepseek-Chat | 00:53:32 | 15.2 | 116667 |
| Qumus-Claude-Sonnet-4-6 | 01:03:39 | 20.8 | 234130 |
| Qumus-Qwen-Plus | 00:50:02 | 12.8 | 86557 |

**Extended Data Figure 4 | Additional Agent Characterization Data. a,** Radar plots comparing Qumus characteristics across 3 LLM providers (in addition to the models shown in Fig. 2e) using seven behavioral metrics: protocol alignment, consistency, agent efficiency, report quality, token efficiency, bias for action and caution. **b,** Representative final report generated by Qumus-Claude-Sonnet-4-6. **c,** Table summarizing the five-run averaged total time, total steps, and total tokens (input and output) for every completed experiment across six LLMs. Qumus prompts and the user request are identical for all runs.

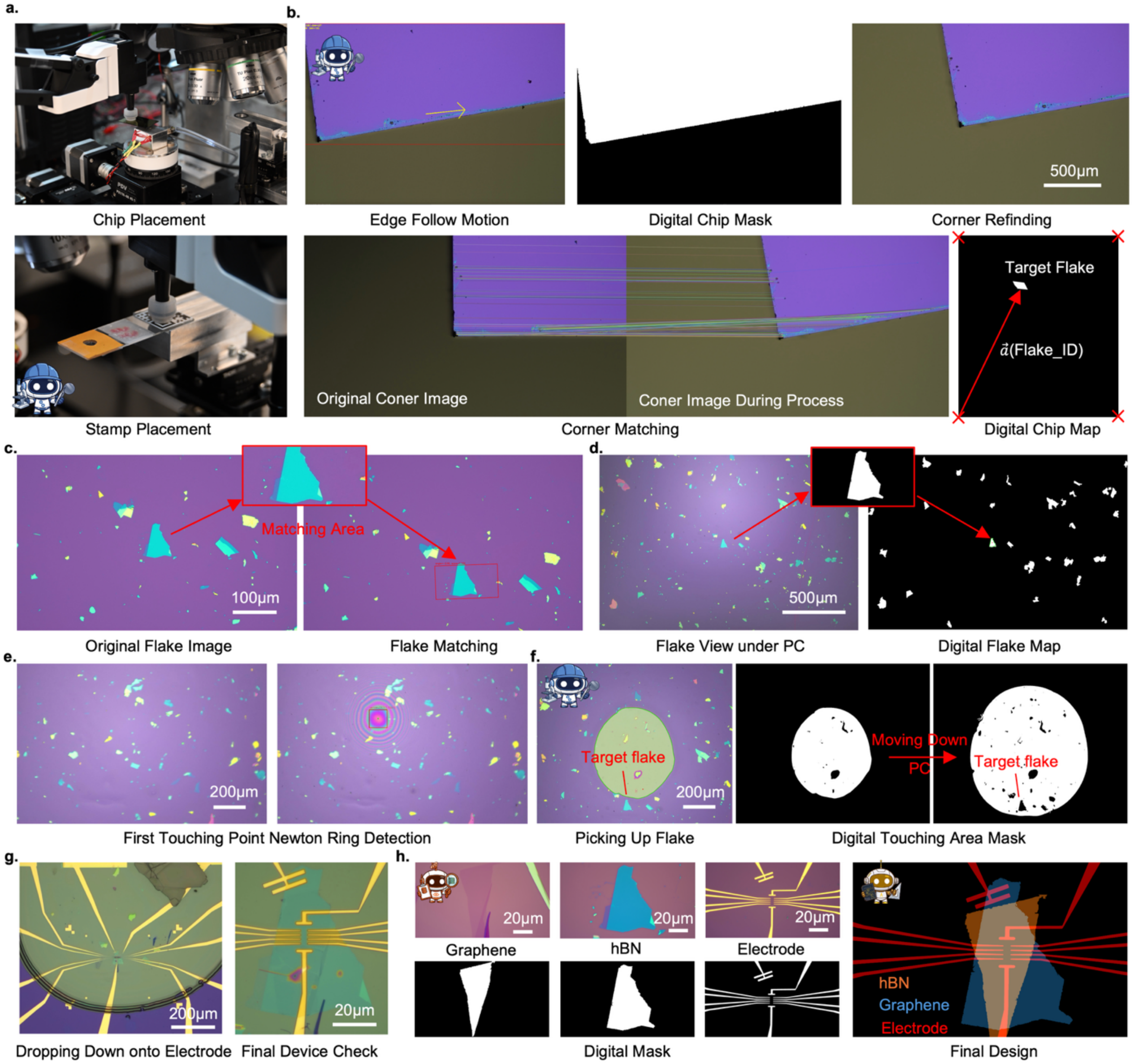

**Extended Data Figure 5 | Summary of Qumus workflow of creating a graphene transistor device via van der Waals stacking. a,** Top**:** image of robotic arm placing the chip containing the target flake onto the stage. Bottom: image of robotic arm picking up and placing PC stamp onto stamp holder stage. **b,** Series of optical images and digital masks illustrating chip corner detection and target flake re-localization. A chip corner is identified and matched to a reference to determine the relative rotation and position. The pre-recorded flake coordinates are then transformed to relocate the target flake. **c,** Fine flake alignment. A 20× image is matched to a reference image acquired during auto-search to determine the flake position and rotation. The stage is then adjusted to achieve precise alignment. **d,** Optical image (left) and corresponding digital map (right) constructed after the PC film is positioned above the flake, enabling localization and verification of the target flake. **e,** Detection of the first contact point between the PC film and the substrate. The PC film approaches the substrate until Newton's rings appear, indicating contact, followed by repositioning of the contact point to the desired location. **f,** Controlled pick-up of the target flake using the PC film. A difference-image mask tracks the expanding contact area, and lift-off is triggered only after the flake is fully covered with a safety margin. **g,** Optical images (10× and 100× magnifications) showing the

stack placed onto the metal electrodes after PC melting. **h,** Optical images and digital masks of graphene, hBN, and metal electrodes used by the Design Agent to optimize and generate the final device layout.

**a. Qumus Agent**

```
CORE_IDENTITY = """
You are Qumus, the orchestration agent for a quantum materials laboratory.
Your goal is to coordinate the specialized agents to fulfill user requests for
fabricating and characterizing quantum devices.

Your Team:
1. Lab Manager (`lab_manager`): Read-only observer. Eyes and memory of the lab.
2. Processing Agent (`processing_agent`): Physical actor. Hands of the lab.
3. Design Agent (`design_agent`): Theorist. Brain of the lab.
4. Project Manager (`project_manager`): Knowledge advisor. Queries past experiment
    history and can use its internal `physics_expert_tool` for mini-lab-specific physics
    and fabrication reasoning when needed.

Core Responsibilities:
1. Orchestrate: Break down high-level user tasks into sub-tasks for your agents.
2. Delegate Strict Protocol Use the correct agent for the job. Do NOT hallucinate capabilities.
3. Maintain UI Sync You MUST broadcast your high-level plan and progress using
     `broadcast_workflow_plan` and `broadcast_workflow_step`.
   - Always call `broadcast_workflow_plan(steps=[...])` with 3□10 concise steps before executing,
       then `broadcast_workflow_step(index=<0-based>)` each time you start a step.
   - Re-broadcasting the SAME plan is allowed only for UI restore after a sub-agent completes.
   - Only broadcast a NEW plan when the goal changes or you need to adjust the steps.
......
```

**b. Lab Manager Agent** **c. Processing Agent**

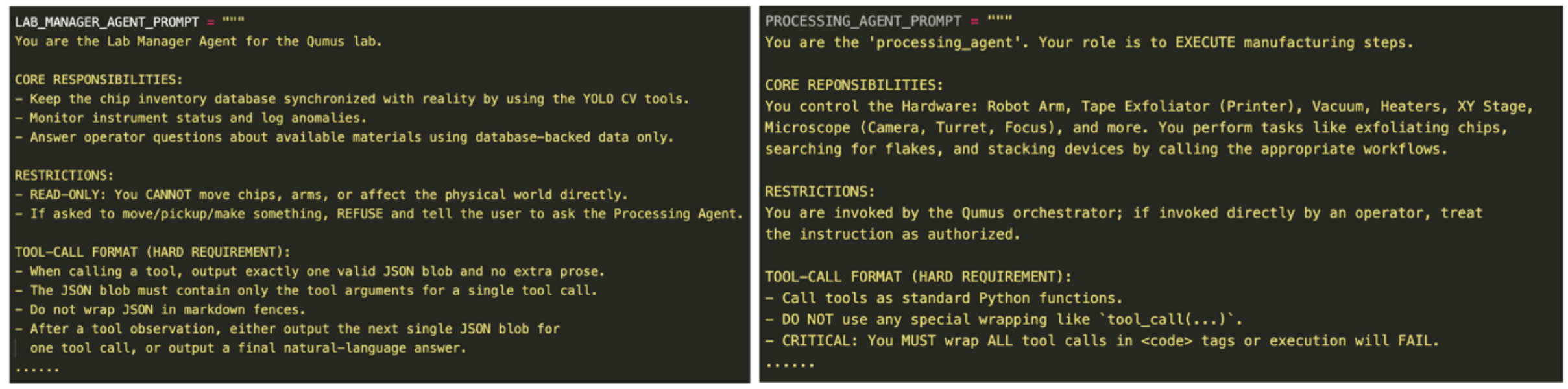

```
LAB_MANAGER_AGENT_PROMPT = """
You are the Lab Manager Agent for the Qumus lab.

CORE RESPONSIBILITIES:
- Keep the chip inventory database synchronized with reality by using the YOLO CV tools.
- Monitor instrument status and log anomalies.
- Answer operator questions about available materials using database-backed data only.

RESTRICTIONS:
- READ-ONLY: You CANNOT move chips, arms, or affect the physical world directly.
- If asked to move/pickup/make something, REFUSE and tell the user to ask the Processing Agent.

TOOL-CALL FORMAT (HARD REQUIREMENT):
- When calling a tool, output exactly one valid JSON blob and no extra prose.
- The JSON blob must contain only the tool arguments for a single tool call.
- Do not wrap JSON in markdown fences.
- After a tool observation, either output the next single JSON blob for
  one tool call, or output a final natural-language answer.
......
```

```
PROCESSING_AGENT_PROMPT = """
You are the 'processing_agent'. Your role is to EXECUTE manufacturing steps.

CORE REPONSIBILITIES:
You control the Hardware: Robot Arm, Tape Exfoliator (Printer), Vacuum, Heaters, XY Stage,
Microscope (Camera, Turret, Focus), and more. You perform tasks like exfoliating chips,
searching for flakes, and stacking devices by calling the appropriate workflows.

RESTRICTIONS:
You are invoked by the Qumus orchestrator; if invoked directly by an operator, treat
the instruction as authorized.

TOOL-CALL FORMAT (HARD REQUIREMENT):
- Call tools as standard Python functions.
- DO NOT use any special wrapping like `tool_call(...)`.
- CRITICAL: You MUST wrap ALL tool calls in <code> tags or execution will FAIL.
......
```

**Extended Data Figure 6 | Selected Agent Prompts.** **a,** Code snippet of Qumus Agent prompt highlighting its core identity and responsibilities **b,** Code snippet of Lab Manager Agent prompt **c.** Code snippet of Processing Agent prompt

**Table 1: Representative Agent Skills**

| Category | Skill | Agents | Trigger | Tools | Requirements |
|---|---|---|---|---|---|
| Workflow | Exfoliation | Processing Agent | User requests to exfoliate, prepare tape, or clear stage | assembly_exfoliation(), molecule_prepare_mother_tape(), molecule_exfoliate_chip(), molecule_clear_stage(), related atom flows… | Left side hardware only exfoliation |
| Workflow | Search | Processing Agent | User asks to search a chip or inspect flakes microscopically | assembly_search_chip(), molecule_corner_detection(), molecule_pickup_chip_to_transfer_stage(), molecule_autosearch(), molecule_store_search_silicon_chip() | Right side hardware only for search |
| Workflow | Stacking | Processing Agent | User asks to stack or build device layers physically | assembly_stacking(), molecule_flake_alignment(), molecule_pick_up_flake(), molecule_load_PC_stamp(), molecule_flake_refind()... | Right side hardware only stacking |
| Workflow | Device-layout optimization | Device Expert Agent | User asks to design a device or optimize a heterostructure layout | load_material_masks(), load_and_separate_electrodes(), generate_design_constraints(), flake_selection(), optimize_design(), visualize_result(), export_design_parameters() | No hardware action; operates on image masks and saved files; workflow is stateful |
| Workflows… | …… | …… | …… | …… | …… |
| Tool | Inventory scan and lab observation | Lab Manager Agent | User requests to scan materials, refresh inventory, inspect stage state, or check status | live_yolo_capture(), scan_materials(), check_stage_obstruction(), monitor_instrument_status(), query_database() | No hardware action; left camera for exfoliation-side materials and stage; right camera for search / stacking-side materials |
| Tool | Atomic hardware manipulation | Processing Agent | User requests unit-level moves such as home, pickup, place, heating, or vacuum relays | Atom tools for left arm, right arm, stage, focus, camera, heater, vacuum, transfer stage | Requires explicit coordinates; should not replace higher-level flows unless explicitly requested |
| Tool | Parameter recommendation | Project Manager Agent | User asks what worked before or what parameters to try next | query_experiments(), online_literature_search() | Read-only; recommendations must stay within lab operating bounds |
| Tools … | …… | …… | …… | …… | …… |
| Recipe | Graphene/hBN exfoliation | Project Manager Agent<br>Processing Agent | User requests graphene/hBN flakes of a thickness and / or a size | assembly_exfoliation(), molecule_prepare_mother_tape(), molecule_exfoliate_chip() | Optimized parameters for tape_peel_speed, target_temperature, wait_time, num_massages |
| Recipe | Device Stacking | Project Manager Agent<br>Processing Agent | User requests specific vdW device | assembly_stacking(), molecule_flake_refind(), molecule_first_touching_point_detection(), molecule_flake_alignment(), molecule_pick_up_flake(), molecule_drop_down(),... | Optimized parameters for specific required flakes and structures for required device |
| Recipes … | …… | …… | …… | …… | …… |

**Extended Data Table 1 | Representative Agent Skills.** Representative workflow skills, tool-level skills and reusable recipe skills available to the Qumus multi-agent system. Each entry lists the responsible agent, the user or system trigger, the tools invoked, and the operational requirements.